\documentclass[doublecol]{epl2}
\usepackage{graphicx}

\title{Phase transitions driven by L\'evy stable noise: exact solutions and stability analysis of nonlinear fractional Fokker-Planck equations}
\shorttitle{Phase transitions driven by L\'evy stable noise} 

\author{A. Ichiki\thanks{E-mail: \email{aichiki@mikan.ap.titech.ac.jp}} \and M. Shiino}
\shortauthor{A. Ichiki and M. Shiino}

\institute{                    
  Department of Applied Physics, Faculty of Science, Tokyo Institute of Technology, 2-12-1 Ohokayama Meguro-ku Tokyo, Japan
}

\abstract{
Phase transitions and effects of external noise on many body systems are one of the main topics in physics. In mean field coupled nonlinear dynamical stochastic systems driven by Brownian noise, various types of phase transitions including nonequilibrium ones may appear. A Brownian motion is a special case of L\'evy motion and the stochastic process based on the latter is an alternative choice for studying cooperative phenomena in various fields. Recently, fractional Fokker-Planck equations associated with L\'evy noise have attracted much attention and behaviors of systems with double-well potential subjected to L\'evy noise have been studied intensively. However, most of such studies have resorted to numerical computation. We construct an {\it analytically solvable model} to study the occurrence of phase transitions driven by L\'evy stable noise. }

\begin{document}

\maketitle

\section{Introduction}
Phase transitions in coupled nonlinear dynamical systems are one of the main topics in physics, especially in statistical physics, and have attracted much attention not only in natural sciences but in social sciences, since such studies are required to understand various types of cooperative phenomena in ensembles consisting of elements interacting with each other. 
Especially for systems with mean field couplings, owing to the simplification of interactions, phase transitions including nonequilibrium types of phase transitions that occur as a consequence of ergodicity breaking are investigated in detail and bifurcations of temporal attractors such as limit cycle and chaos with changes in the noise amplitudes have been found \cite{chaos1,chaos2,Aihara} (see also \cite{Hinrichsen}). 
The starting point of such studies is usually a set of Langevin equations driven by white Gaussian noise, more precisely Brownian noise. 
By taking advantage of the self-averaging properties for mean field couplings, the master equation describing the temporal evolution of the empirical probability density of the system becomes a nonlinear Fokker-Planck equation (NFPE) \cite{chaos1,chaos2,ms85,ms87,delay,TDF} in the thermodynamic limit. 
The origin of the nonlinearity in NFPE with respect to the probability density is the mean field coupling term which gives rise to averaging over the empirical probability density in the thermodynamic limit. 
Especially for the nonlinearly coupled stochastic dynamics with quadratic potential, the form of the NFPE with the self-consistently determined mean field coupling term coincides formally with the form of the linear Fokker-Planck equation corresponding to the well-known Ornstein-Uhlenbeck (OU) process, i.e., the OU process of onebody system. 
Since the probability density for the OU process after large times takes a Gaussian form according to an H-theorem, the temporal evolution of the order parameters is analytically and easily obtained for the mean field coupled system with quadratic potential. 

L\'evy stable noise \cite{Levy} is considered to be as important as Brownian noise in dynamical stochastic systems. 
Brownian noise is a special case of L\'evy stable noise, and L\'evy noise distribution exhibits the asymptotic power-law decay whereas Brownian noise distribution decays exponentially. 
Recently anomalous diffusion processes associated with L\'evy stable noise have attracted much attention in a variety of fields not only of natural sciences such as physics, biology, earth science, etc., but of social sciences such as risk management, finance, etc. 
In the context of physics, the microscopic description of the L\'evy processes associated with anomalous diffusion has been proposed in \cite{Lubashevsky}. 
L\'evy flights have been realized by the displacement of the random walker subjected to Gaussian multiplicative noise as a random force. 
In such a system, the statistical property of the displacement of the random walker depends on its "memory" at small time scale, while it obeys fully Markoffian process, which reproduces a L\'evy flight, at large time scale. 
Furthermore, one of interesting applications in physics can be found in subrecoil laser cooling of atoms \cite{laser}, and the studies on this system show the validity of use of L\'evy statistics in physics. 
To study subrecoil laser cooling process, the quantum mechanical analysis on the interactions between atoms and photons, which is significantly difficult in the system of dimension $d\ge 2$, is required. 
On the other hand, dealing with this problem as a classical inhomogeneous random walk (anomalous diffusion process owing to the long lifetimes of atoms with constant momentum close to zero) of 'dressed atoms' in momentum space allows us to use L\'evy statistics to easily obtain the behaviors of cooled atoms. 
This approach is shown to be equivalent to more complicated microscopic quantum approach, and a good agreement between theoretical and experimental results is reported. 
Furthermore, L\'evy noise was studied from the viewpoint of improving neural signal detection \cite{Patel}. 

L\'evy statistics, in general, is expected to be a powerful tool to investigate dynamical stochastic processes driven by long tail distributed noise, the superposition of a sufficiently large number of which {\it does not} follow the central limit theorem. 
Such a superposition is known to obey a L\'evy stable distribution. 
This fact is regarded as the generalization of the central limit theorem. 

While the stationary probability density {\it does not} take a Gibbsian form in the system with L\'evy noise, it is shown that an H-theorem applicable to systems with L\'evy noise holds \cite{Htheorem}. 
Surprisingly, the H-functional for such systems also takes the form of the relative entropy, the same form as the one for the systems with Brownian noise. 
Considering such correspondences in spite of the significantly different statistical behaviors between Brownian noise and L\'evy noise, it will be important to investigate how statistical properties, especially cooperative phenomena in systems with Brownian noise is generalized and extended in systems with L\'evy noise. 
The main target of this paper is to elucidate how behaviors of the nonlinear coupled stochastic dynamics with mean field couplings in the presence of L\'evy noise differ from those in the presence of Brownian noise, in terms of the existence of the equilibrium phase transition, stability condition for equilibrium solution, and the criticality. 
The phrases "bifurcation" or "phase transition" have been used to describe the statistical properties of {\it onebody} stochastic systems in some literatures \cite{Shpyrko, Chechkin, Chechkin2}. 
The "bifurcations" associated with L\'evy noise have also been investigated to understand the difference of the influence on systems between L\'evy and Gaussian noise \cite{Chechkin, Chechkin2}. 
There, however, the "bifurcations" only imply the change of the form of the (stationary) probability density and are {\it not bifurcations in the mathematical sense}. 
The bifurcations presented in this paper are totally different from them. 
We want to be concerned with {\it genuine phase transitions or bifurcations where the stability exchanges between attractors occur}. 
We will present {\it rigorous analyses to obtain exact results} in this paper. 

\section{Model}
We deal with the following stochastic dynamics of $N$ one-dimensional elements: 
\begin{equation}
\label{Nbodydynamics}
\upd x_{j}(t)=m_{j}\left(\left\{ x_{i}(t)\right\} ,t\right)\upd t
+\sigma\left(\left\{ x_{i}(t)\right\} ,t\right)\upd L_{j}\, ,
\end{equation}
where $m$ and $\sigma$ denote functions of $x_{1}(t),\cdots ,x_{N}(t)$ and time $t$, $L_{j}$ is a L\'evy stable motion and $\upd L_{j}$ is determined by its characteristic function involving four parameters ($\alpha$, $\beta$, $\gamma$, $D$): 
$E\left[\exp\left( ik\upd L_{j}\right)\right] 
=
\exp\left(\upd t \left\{
ik\gamma -D|k|^{\alpha} \left[ 1-i\beta\frac{k}{|k|}\omega (k,\alpha )
\right]\right\}\right) + \mathrm{o}(\upd t)$ 
with 
$\omega (k,\alpha )=\tan\frac{\pi\alpha}{2}$ for $\alpha\neq 1$, 
$\omega (k,\alpha )=-\frac{2}{\pi}\ln |k|$ for $\alpha=1$. 
$E\left[\cdot\right]$ denotes the average over the L\'evy noise $\upd L_{j}$. 
Here the parameter $\alpha$ ($0<\alpha \le 2$) characterizes the asymptotic tail of the L\'evy stable distribution $f(L;\alpha ,\beta ,\gamma ,D)$ for $\alpha <2$ as $f(L;\alpha ,\beta ,\gamma ,D)\sim |L|^{-\alpha -1}$ with $|L|\gg 1$. 
For $\alpha =2$, the L\'evy stable distribution reproduces a Gaussian distribution. 
The parameter $\beta$ ($-1\le \beta \le 1$) is the skewness parameter defining the degree of asymmetry of the stable distribution, $\gamma$ ($-\infty <\gamma <\infty$) is the center or location parameter which denotes the mean value of the distribution when $\alpha > 1$ and  $D$ ($0\le D<\infty$) is called the scale parameter which represents the generalized diffusion coefficient. 
The function $\omega (k,\alpha )$ is discontinuous at $\alpha =1$. 
The alternative choice for $\omega (k,\alpha )$ continuous with respect to $\alpha$ has been presented in some literatures \cite{Zolotarev,Dybiec2004}. 
However we will use the standard choice for $\omega (k,\alpha )$ in this paper. 

In this paper we consider that the L\'evy stable noise $\upd L_{j}(t)$'s are white noise and statistically independent with respect to the site $j$, i.e., they follow i.i.d. 

For the purpose of dealing with an analytically solvable model, hereafter the parameters $\alpha$, $\beta$, $\gamma$ and $D$ are assumed to take same value for every site $j$ and we treat the case of 
\begin{equation}
\label{mj}
m_{j}\left( x(t),t\right)=-ax_{j}(t)+\frac{J}{N}\displaystyle\sum_{i=1}^{N}
F\left( x_{i}(t)\right)
\end{equation}
with $a$ positive constant, $J$ constant, $F(x)$ an arbitrary bounded nonlinear function and $\sigma$ in eq.~(\ref{Nbodydynamics}) a positive constant. 
Eq.~(\ref{mj}) is well known to describe analog neural network models with $F$ representing a transfer function, which is usually taken to be a bounded function such as a sigmoid function. 
A choice of non-monotonic functions for $F$ was shown to improve the performance of associative memory neural networks \cite{ShiinoFukai2003PRE}. 
It is also noted that the linearlity of $m_{j}$ in the variable $x_{j}$ makes the model analytically tractable \cite{chaos2} as shown below. 
In this model, since the self-averaging property or the law of large numbers will hold for the mean field coupling term $M(t)\equiv\frac{1}{N}\sum_{j}F\left( x_{j}(t)\right)$ in the large $N$ limit, the dynamics of the $N$-body system (\ref{Nbodydynamics}) is expected to be reduced to dealing with the onebody stochastic dynamics: 
\begin{eqnarray}
\label{onebodydynamics}
\upd x(t)&=&m\left( x(t),t\right)\upd t + \sigma\upd L\,,\nonumber\\
m\left( x(t),t\right)&=&-ax(t)+JM(t)\, .
\end{eqnarray}
Notice that the dynamics (\ref{onebodydynamics}) in the case of $\alpha =2$, i.e., in the special case where the stochastic process $L$ is a Brownian motion, becomes the well-known OU process. 
This type of onebody stochastic dynamics driven by L\'evy stable noise (\ref{onebodydynamics}) without the mean field interaction term $M$ has been investigated in \cite{Dybiec}. 
The important difference between the model presented in \cite{Dybiec} and our model is the presence of the coupling term $M$ in our model, which makes the time evolution equation of the probability density $p(x,t)$ nonlinear owing to $M(t)=\int_{-\infty}^{\infty}F(x)p(x,t)\upd x$. 
Hence phase transitions may appear in our model. 

\section{Fractional Fokker-Planck equation and stationary solutions}
\begin{figure}
\onefigure[scale=0.65]{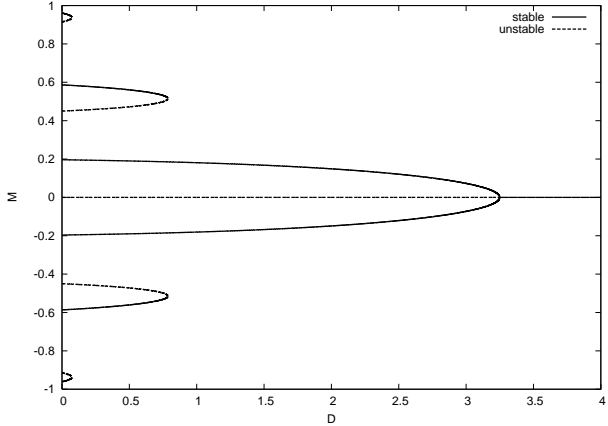}
\caption{Solution of the order parameter equation with $F(x)=\sin x$, $J=15$, $a=1$, $\sigma =1$, $\alpha =1.2$,  $\gamma =0$, $\beta =0$. 
The stable solution is plotted in the solid line and the unstable is in the broken line. 
The critical value of the generalized diffusion coefficient is evaluated analytically as $D_{\mathrm{c}}=\left( a\alpha/\sigma^{\alpha}\right)\ln (J/a)\sim 3.25$. }
\label{fig1}
\end{figure}

The FPE corresponding to the stochastic differential equation with L\'evy noise takes the form of fractional FPE \cite{Htheorem,ext,char1,char2}. 
According to \cite{char1,char2}, the nonlinear fractional FPE (NFFPE) associated with the stochastic dynamics (\ref{onebodydynamics}) will be given as 
\begin{eqnarray}
\label{FPE}
&&\frac{\partial p(x,t|x_{0},t_{0})}{\partial t}=
-\frac{\partial}{\partial x}\left[ m(x,t)
+\gamma\sigma\right] p(x,t|x_{0},t_{0})\nonumber\\
&+&\mathcal{F}^{-1}\left( -D\sigma^{\alpha}|k|^{\alpha}\left[ 1-i\beta
\frac{k}{|k|}\omega (k\sigma ,\alpha )\right]\right) [x]\nonumber\\
& \ast & p(x,t|x_{0},t_{0})\, ,
\end{eqnarray}
where $\mathcal{F}^{-1}(\cdot )$ denotes the inverse Fourier transform and $\ast$ represents the convolution. 
Using the Fourier transform of the probability density $\hat{p}(k,t)=\int_{-\infty}^{\infty}p(x,t)\exp (ikx)\, \upd x$, the FPE (\ref{FPE}) is rewritten as 
\begin{eqnarray}
\label{FPE2}
\frac{\partial \hat{p}}{\partial t}&=&-ak\frac{\partial \hat{p}}{\partial k}
+ik(JM+\gamma\sigma )\hat{p}\nonumber\\
&-&D\sigma^{\alpha}|k|^{\alpha}
\left[ 1-i\beta\frac{k}{|k|}\omega (k\sigma ,\alpha )\right]\hat{p}\, .
\end{eqnarray}
In the case of $\alpha\neq 1$, one can straightforwardly show that the temporal solution for the FPE (\ref{FPE2}) {\it after large times}  takes the form 
$\hat{p}(k,t)=\exp\left[ i\gamma_{\mathrm{eff}}k-D_{\mathrm{eff}}|k|^{\alpha}\left( 1-
i\beta_{\mathrm{eff}}\frac{k}{|k|}\tan\frac{\pi\alpha}{2}\right)\right]$
and the parameters $\beta_{\mathrm{eff}}$, $\gamma_{\mathrm{eff}}$ and $D_{\mathrm{eff}}$ obey the dynamics: 
$\frac{\upd \gamma_{\mathrm{eff}}}{\upd t}=-a\gamma_{\mathrm{eff}}+JM+\gamma\sigma$, 
$\frac{\upd D_{\mathrm{eff}}}{\upd t}=-a\alpha D_{\mathrm{eff}}+D\sigma^{\alpha}$, 
$\frac{\upd}{\upd t}\left(\beta_{\mathrm{eff}}D_{\mathrm{eff}}\right)=
-a\alpha D_{\mathrm{eff}}\beta_{\mathrm{eff}}+D\beta \sigma^{\alpha}$. 
In principle, $M(t)$ is evaluated in terms of $\beta_{\mathrm{eff}}$, $\gamma_{\mathrm{eff}}$ and $D_{\mathrm{eff}}$ by the definition of $M(t)$ and the above-mentioned temporal solution $\hat{p}(k,t)$. 
For some special bounded functions $F(x)$, the concrete form of $M(t)$ can be represented {\it explicitly} in terms of an analytic function of these parameters. 
In particular, for $F(x)=\sin x$, assuming $\beta =\gamma =0$ and the initial condition $p(x,t=0)=\delta (x-x_{0})$, the temporal solution (for any $t$) is given by solving eq.~(\ref{FPE2}) as 
\begin{eqnarray}
\hat{p}(k,t)&=&\exp\left\{ -\frac{D\sigma^{\alpha}|k|^{\alpha}}{a\alpha}\left[ 1-\exp (-a\alpha t)\right] +ikB(t)\right\}\,,\nonumber\\
B(t)&=&x_{0}\exp (-at)
+\int_{0}^{t}\upd s\left[ JM(s)\right]\exp\left[ -a(t-s)\right]\,.\nonumber\\
&&
\end{eqnarray} 
Then the time evolution of the order parameter $M(t)$ is given by solving $M(t)=\exp\left[ -D\sigma^{\alpha}\left( 1-e^{-a\alpha t}\right) /(a\alpha )\right]\sin B(t)$ together with 
\begin{equation}
\label{Bt}
\frac{\upd B(t)}{\upd t} + aB(t)=J\exp\left[ -\frac{D\sigma^{\alpha}\left( 1-e^{-a\alpha t}\right)}{a\alpha }\right] \sin B(t)
\end{equation}
with $B(0)=x_{0}$ [Shiino M. and Ichiki A., $8^{th} ${\it Asia-Pacific Conference of Fundamental Problems of Opto and Microelectronics} (Kokushikan University, Tokyo) 2008]. 
Hereafter we will confine ourselves only to the case of $\alpha\neq 1$ for simplicity. 
Then the stationary solution of (\ref{FPE2}) is obtained as 
\begin{eqnarray}
\label{equil}
\hat{p}_{\mathrm{eq}}(k)&=&\exp\Bigg[ \frac{i(JM+\gamma\sigma)}{a}k \nonumber\\
&-& \frac{D|k|^{\alpha}\sigma^{\alpha}}{a\alpha}\left( 1-i\beta
\frac{k}{|k|}\tan\frac{\pi\alpha}{2}\right)\Bigg]\, .
\end{eqnarray}
In general, the moments related to the stationary solution (\ref{equil}) diverge. 
However, the mean field interaction term $M$ is finite since $F(x)$ is bounded. 
It is worth to mention that the temporal solution $\hat{p}(k,t)$ after large times corresponds to L\'evy stable distribution. 
This is reminiscent of a similar situation described by the H-theorem for the stochastic processes with Brownian noise presented in \cite{chaos1,chaos2}, which ensures the convergence to a temporal probability density of Gaussian form after large times. 
The proof of the convergence to the L\'evy stable distribution in our model will be presented in detail elsewhere. 
Notice that the solution (\ref{equil}) reproduces the equilibrium probability density for the usual OU process when $\alpha =2$. 
Using (\ref{equil}), the stationary value $M_{0}$ of the temporal order parameter $M(t)$ is determined self-consistently as the solution of the order parameter equation 
$M_{0}=\exp\left(\frac{-D\sigma^{\alpha}}{a\alpha}\right)\sin\left[
\frac{JM_{0}+\gamma\sigma}{a}+\frac{D\sigma^{\alpha}\beta \tan (\pi\alpha /2)}{a\alpha}\right]$ 
in the case of $F(x)=\sin x$. 
The solution of this order parameter equation is plotted as a function of the generalized diffusion coefficient $D$ in fig.~\ref{fig1}. 
We can see that phase transitions occur and stability of the order parameter changes with a change in the parameter $D$. 
The value of $D$ when the nontrivial stable solution $M_{0}\neq 0$ vanishes with a gradual increase in $D$ is analytically evaluated as $D_{\mathrm{c}}=\left( a\alpha\right)\ln\left( J/a\right)$ in the case of $F(x)=\sin x$, $\alpha\neq 1$, $\beta =\gamma =0$, $\sigma =1$. 
In general, the critical value $D_{\mathrm{c}}$, which is the value of $D$ when the stability of $M_{0}$ alters, is evaluated by solving the equation 
$J\int p_{\mathrm{eq}}(x)\left[\upd F(x)/\upd x\right]\upd x=a$
with respect to $D$ according to the stability analysis for the stationary solutions mentioned bellow. 
Furthermore, it is found that the stable solution $M_{0}\neq 0$ scales as $M_{0}\propto\left(D_{\mathrm{c}}-D\right)^{1/2}$
near the critical point $D<D_{\mathrm{c}}$ in accordance with the mean-field universality class, by using the stationary solution (\ref{equil}) and expanding the definition of the order parameter $M_{0}$ with respect to the difference in the generalized diffusion coefficient $D_{\mathrm{c}}-D$. 
Notice that this scaling law also holds in the Brownian noise case since this law is independent of the parameter $\alpha$. 

\section{Stability analysis}
In what follows, we investigate the stability of the stationary solution $M_{0}$. 
Considering a small deviation of the probability density from the stationary density $\delta p(x,t)\equiv p(x,t)-p_{\mathrm{eq}}(x)$, the Fourier transform of this deviation turns out to follow the FPE 
\begin{eqnarray}
\label{delFPE}
\frac{\partial\delta\hat{p}}{\partial t}&=&-ak\frac{\partial\delta\hat{p}}{\partial k}
+ik(JM_{0}+\gamma\sigma )\delta\hat{p}+ikJ\delta M\hat{p}_{\mathrm{eq}}\nonumber\\
&-&D\sigma^{\alpha}|k|^{\alpha}\left(
1-i\beta\frac{k}{|k|}\tan\frac{\pi\alpha}{2}\right)\delta\hat{p}
+\mathcal{O}(\delta^{2})\, ,
\end{eqnarray}
where $\delta M$ is a small deviation of the order parameter $M$ related to the deviation of the probability density $\delta p$: $\delta M(t)=\int F(x)\delta p(x,t)\upd x$. 
Here we consider the eigenfunction of the Fokker-Planck operator in eq.~(\ref{delFPE}), i.e., $\frac{\partial\delta\hat{p}_{\lambda}}{\partial t}=\lambda\delta\hat{p}_{\lambda}$. 
Such a solution of the FPE (\ref{delFPE}) is 
\begin{eqnarray}
\label{eigen}
\delta\hat{p}_{\lambda}(k,t)&=&\left(\frac{iJk}{a+\lambda}+
\frac{1-\frac{J}{a+\lambda}\langle F^{\prime}\rangle}{R}|k|^{-\lambda /a}\right)\nonumber\\
&\times&
\delta M(t=0)\exp\left( \lambda t\right)\hat{p}_{\mathrm{eq}}(k)\, ,
\end{eqnarray}
where $\delta M(t=0)$ the deviation of the order parameter $M$ from the stationary solution $M_{0}$ at time $t=0$ and 
$\langle F^{\prime}\rangle\equiv\int \frac{\upd F(x)}{\upd x}p_{\mathrm{eq}}(x)\upd x$, 
$R\equiv\int\upd x\int\frac{\upd k}{2\pi}F(x)|k|^{-\lambda /a}\hat{p}_{\mathrm{eq}}(k)\exp\left( -ikx\right)$. 
Since $\delta\hat{p}_{\lambda}(k=0,t)=0$ for normalization condition, $1-J\langle F^{\prime}\rangle /(a+\lambda )=0$ is required for the case $\lambda /a > 0$. 
Since we assume $a>0$ for physical interest, the positive eigenvalue of the Fokker-Planck operator $\lambda$ is given as $\lambda = -a+J\langle F^{\prime}\rangle$.
Hence the condition for stable stationary solution $M_{0}$ is 
\begin{equation}
\label{stabilitycondition}
a-J\langle F^{\prime}\rangle > 0\,.
\end{equation}
Notice here that the $\alpha$-dependence of the stability condition (\ref{stabilitycondition}) exists only in the quantity $\langle F^{\prime}\rangle$, and this result reproduces the stability condition in the Brownian noise case when $\alpha =2$. 
For the choice of $F(x)=\sin x$, one finds $R=M_{0}$, $\langle F^{\prime}\rangle =\frac{1}{2}\left[\hat{p}_{\mathrm{eq}}(1)+\hat{p}_{\mathrm{eq}}(-1)\right]$ and then obtains the stability condition: 
$0 < a-J\exp\left( -\frac{D\sigma^{\alpha}}{a\alpha}\right)\cos\left[
\frac{JM_{0}+\gamma\sigma}{a}+\frac{\beta\sigma^{\alpha} D\tan (\pi\alpha /2)}{a\alpha}\right]$. 
The stable and unstable stationary solutions in this case are plotted in fig.~\ref{fig1}. 

\section{Summary}
In summary, we have investigated how the properties of the nonlinear coupled stochastic dynamics with mean field couplings in the presence of Brownian noise is modified in the presence of L\'evy noise. 
We have shown the occurrence of bifurcations for mean field type NFFPE associated with the stochastic process which describes nonlinearly coupled elements subjected to independent identical L\'evy noise. 
All of the analyses have been conducted {\it rigorously} in the thermodynamic limit for systems under the influence of L\'evy noise with index $\alpha\neq 1$. 
Similar analysis can be carried out also in the case of $\alpha =1$. 
The detailed analysis including the case of $\alpha =1$ will be reported elsewhere. 
In this paper, the temporal evolution of the order parameter has been obtained together with the stability condition for its stationary value. 
We have also presented the scaling of the order parameter near the critical point. 
The exact results presented in this paper have been obtained by taking advantage of (i) the self-averaging property, or the law of large numbers owing to the presence of mean field couplings in the thermodynamic limit, (ii) the identical dynamics for every element constituting the system and (iii) the quadratic potential in the intrinsic dynamics. 
The choice of such potential gives rise to a linear system for which L\'evy distribution holds. 
The situation is reminiscent of the so called OU process with a Gaussian probability density that manifests itself in a linear system subjected to Brownian white noise. 
The case with a double-well potential will be of interest and is now under way. 
There also remains open problems of how fluctuation-dissipation theorems and critical fluctuations are modified under the condition that moments of L\'evy distribution in general do not exist. 

In literatures \cite{Chechkin, Chechkin2}, the terminology "bifurcation" is used to describe the situation where the form of the probability density of a single element subjected to a L\'evy noise changes between unimodal and bimodal functions. 
On the other hand, in this paper, {\it genuine phase transitions or bifurcations driven by L\'evy noise have been shown}. 
Since in recent years the effects of L\'evy noise have been extensively studied in various fields of science, a study of fundamental problems of phase transitions involving L\'evy noise should be of vital importance. 
Especially a rigorous analysis of such cooperative phenomena based on analytically tractable models is quite useful considering potentially wide applications of it. 
Detailed analyses including the derivation of eq.~(\ref{Bt}) and related issues as well as the case $\alpha =1$ will be reported elsewhere.

\acknowledgments
One of the authors (A. I.) is supported by the Grant-in-Aid for JSPS Fellows No. 20.9513.


\begin{thebibliography}{0}

\bibitem{chaos1}
\Name{Shiino M. \and Yoshida K.}
\REVIEW{Phys. Rev. E}{63}{2001}{26210}.

\bibitem{chaos2}
\Name{Ichiki A., Ito H. \and Shiino M.}
\REVIEW{Physica E}{40}{2007}{402}.

\bibitem{Aihara}
\Name{Kanamaru T. \and Aihara K.}
\REVIEW{Neural Computation}{20}{2008}{1951}.

\bibitem{Hinrichsen}
\Name{Hinrichsen H.}
\REVIEW{J. Stat. Mech.: Theor. Exp.}{}{2007}{P07006}.

\bibitem{ms85}
\Name{Shiino M.}
\REVIEW{Phys. Lett. A}{112}{1985}{302}.

\bibitem{ms87}
\Name{Shiino M.}
\REVIEW{Phys. Rev. A}{36}{1987}{2393}.

\bibitem{delay}
\Name{Shiino M. \and Doi K.}
\Book{Proceedings of the 2007 IEEE Symposium on Foundations of Computational Intelligence} 
\Publ{IEEE Press, NJ}
\Year{2007}
\Page{100}.

\bibitem{TDF}
\Name{Frank T. D.}
\Book{Nonlinear Fokker-Plank equations Fundamentals and Applications}
\Publ{Springer-Verlag, Berlin}
\Year{2005}.

\bibitem{Levy}
\Name{L\'evy P.}
\Book{Theorie de l'Addition des Variables Aleatoires}
\Publ{Gauthier-Villars, Paris}
\Year{1937}.

\bibitem{Lubashevsky}
\Name{Lubashevsky I., Friedrich R. \and Heuer A.}
\REVIEW{Phys. Rev. E}{79}{2009}{011110}.

\bibitem{laser}
\Name{Bardou F., Bouchaud J. P., Aspect A. \and Tannoudji C. C.}
\Book{L\'evy Statistics and Laser Cooling}
\Publ{Cambridge University Press, UK}
\Year{2002}.

\bibitem{Patel}
\Name{Patel A. \and Kosko B.}
\Book{Proceedings of the 2007 International Conference on Acoustics, Speech and Signal Processing}
\Vol{3}
\Publ{IEEE Press, NJ}
\Year{2007}
\Page{III-1413}

\bibitem{Htheorem}
\Name{Vlad M. O., Ross J. \and Schneider F. W.}
\REVIEW{Phys. Rev. E}{62}{2000}{1743}.

\bibitem{Shpyrko}
\Name{Shpyrko S. \and Ryazanov V. V.}
\REVIEW{Eur. Phys. J. B}{54}{2006}{345}.

\bibitem{Chechkin}
\Name{Chechkin A. V., Klafter J., Gonchar V. Y., Metzler R. \and Tanatarov L. V.}
\REVIEW{Phys. Rev. E}{67}{2003}{010102(R)}.

\bibitem{Chechkin2}
\Name{Chechkin A. V., Gonchar V. Y., Klafter J. \and Metzler R.}
\Book{Advaces in Chemical Physics}
\Editor{Kalmykov Y. P., Coffey W. T. \and Rice S. A.}
\Vol{133}
\Publ{Wiley, NJ}
\Year{2006}
\Page{439}.

\bibitem{Zolotarev}
\Name{Zolotarev V. M.}
\Book{One-dimensional Stable Distributions}
\Publ{American Mathematical Society, Providence}
\Year{1986}.

\bibitem{Dybiec2004}
\Name{Dybiec B. \and Gudowska-Nowak E.}
\REVIEW{Phys. Rev. E}{69}{2004}{016105}.

\bibitem{ShiinoFukai2003PRE}
\Name{Shiino M. \and Fukai T.}
\REVIEW{Phys. Rev. E}{48}{1993}{867}.

\bibitem{Dybiec}
\Name{Dybiec B., Gudowska-Nowak E. \and Sokolov I. M.}
\REVIEW{Phys. Rev. E}{76}{2007}{041122}.

\bibitem{ext}
\Name{Jespersen S., Metzler R. \and Fogedby H. C.}
\REVIEW{Phys. Rev. E}{59}{1999}{2736}.

\bibitem{char1}
\Name{Yanovsky V. V., Chechkin A. V., Schertzer D. \and Tur A. V.}
\REVIEW{Physica A}{282}{2000}{13}.

\bibitem{char2}
\Name{Schertzer D., Larchev\^eque M., Duan J., Yanovsky V. V. \and Lovejoy S.}
\REVIEW{J. Math. Phys: Math. Gen.}{42}{2001}{200}.

\end{thebibliography}
\end{document}